\documentclass[12pt]{article}
\usepackage{graphicx,color,epsfig,rotate}

\newcommand{\be}{\begin{equation}}
\newcommand{\ee}{\end{equation}}
\newcommand{\bea}{\begin{eqnarray}}
\newcommand{\eea}{\end{eqnarray}}
\newcommand{\la}{\langle}
\newcommand{\ra}{\rangle}

\newcommand{\ld}{\left(}
\newcommand{\rd}{\right)}
\newcommand{\lb}{\left\{}
\newcommand{\rb}{\right\}}
\newcommand{\lbr}{\left[}
\newcommand{\rbr}{\right]}

\newcommand{\non}{\nonumber \\ }

\begin{document}

\title{Momentum distribution of itinerant electrons 
in the one-dimensional Falicov-Kimball model}
\author{Pavol Farka\v sovsk\'y\\
Institute  of  Experimental  Physics,  Slovak   Academy   of
Sciences\\
Watsonova 47, 043 53 Ko\v {s}ice, Slovakia}
\date{}
\baselineskip=24pt
\maketitle

\begin{abstract}

The momentum distribution $n_k$ of itinerant electrons in the one-dimensional 
Falicov-Kimball model is calculated for various ground-state
phases. In particular, we examine the periodic phases with period two, 
three and four (that are ground-states for all Coulomb interactions) as well 
as the phase separated states (that are ground states for small Coulomb 
interactions).  For all periodic phases examined the momentum distribution 
is a smooth function of $k$ with no sign of any discontinuity or singular 
behavior at the Fermi surface $k=k_F$. An unusual behavior of $n_k$ 
(a local maximum) is found at $k=3k_F$ for electron concentrations outside 
half-filling. For the phase separated ground states the momentum distribution 
$n_k$ exhibits discontinuity at $k=k_0 < k_F$. This behavior is interpreted
in terms of a Fermi liquid.  

\end{abstract}



\vspace{5mm}

\newpage
\section{Introduction}

Since its introduction in 1969 the Falicov-Kimball model~\cite{Falicov}
has become an important standard model for a description of
correlated fermions on the lattice. The model describes
a two-band system of localized $f$ electrons and 
itinerant $d$ electrons with the short-ranged $f$-$d$ Coulomb
interaction $U$. The Hamiltonian is 

\begin{equation}
H=\sum_{ij}t_{ij}d^+_id_j+U\sum_if^+_if_id^+_id_i+E_f\sum_if^+_if_i,
\end{equation}
where $f^+_i$, $f_i$ are the creation and annihilation 
operators  for an electron in  the localized state at 
lattice site $i$ with binding energy $E_f$ and $d^+_i$,
$d_i$ are the creation and annihilation operators 
for an electron in the conduction band. The conduction 
band is generated by the hopping 
matrix elements $t_{ij}$, which describe intersite
transitions between the sites $i$ and $j$ (as usually, we assume 
that $t_{ij}=-t$ if $i$ and $j$ are nearest neighbors and 
$t_{ij}=0$ otherwise, and all energies are measured in units of $t$).

The model has been used in the literature to study a great variety of 
many-body effects in metals, of which valence and metal-insulator 
transitions, charge-density waves and electronic ferroelectricity are 
the most common examples~\cite{Chomski,Sham}.
It has been applied to a variety of lattices,
one~\cite{Freericks,Gruber1},
two~\cite{Brandt_Schmidt,Kennedy,Gruber2},
three~\cite{Ramirez},
and infinite dimensional~\cite{Brandt_Mielsch}, and occasionally to small
clusters~\cite{Fark1,Fark2,Fark3}. Exact results are available in very few
instances~\cite{Brandt_Mielsch,Lyzwa,Lemb,Gruber3} and general theorems
have been proved for special cases~\cite{Kennedy}.
In spite of the existence of an analytic solution
in $d=\infty$ dimension~\cite{Brandt_Mielsch,Freericks2} and an 
impressive research
activity in the past, the properties of this seemingly simple model are
far from being understood. Indeed, while the ground-state properties 
of the $f$-electron subsystem has been satisfactory explained, only a few
exact results are known concerning the ground-state correlations of the 
itinerant electrons~\cite{itinerant}. Even, some important quantities such as 
a momentum distribution of itinerant electrons has not been explored yet.  
The first attempt to describe ground-state correlations of itinerant 
electrons has been performed recently by Jedrzejewski et al.~\cite{Jedr}.
They calculated a number of one and two-point correlation functions
in order to describe  short and long-range correlations between 
itinerant electrons, as well as  between itinerant and localized 
electrons for various ground states. Calculations have been done
analytically and by means of well-controlled numerical procedures.
In this paper we use the same method to calculate a momentum 
distribution of itinerant electrons for physically the most 
interesting cases. In particular, we examine the periodic phases
with period two, three and four as well as the phase separated 
states that are ground states for small (large) $f$-electron 
concentrations.

\section{The method}
Before showing results for the momentum distribution of itinerant electrons 
in the one-dimensional Falicov-Kimball model let us briefly summarize
the main steps of the calculational method~\cite{Jedr}. 
Since in the spinless version of the Falicov-Kimball model
without hybridization  the $f$-electron occupation
number $f^+_if_i$ of each site $i$ commutes with
the Hamiltonian (1), the $f$-electron occupation number
is a good quantum number, taking only two values: $w_i=1$
or 0, according to whether or not the site $i$ is occupied
by the localized $f$ electron and the Hamiltonian (1) can be written as

\begin{equation}
H=\sum_{ij}t_{ij}d^+_id_j+U\sum_iw_id^+_id_i+E_fN_f,
\end{equation}
where $N_f=\sum_iw_i$ denotes the number of $f$ electrons.

Thus for a given $f$-electron configuration
~$w=\{w_1,w_2 \dots w_L\}$ defined on a one-di\-men\-sional lattice 
(of $L$ sites) with periodic boundary conditions, the Hamiltonian (2)
is the second-quantized version of the single-particle
Hamiltonian $h(w)$. If we denote by $\{|i\ra \}_{i=1,\ldots,L}$ 
the orthogonal basis of one-electron
states, such that $d^+_i$ creates an electron in the state $|i\ra$,
then the matrix elements of $h(w)$ in the basis
$\{ |i \ra \}_{i=1,\ldots,L}$ are defined by
\be
\label{Hw}
H(w)=\sum_{i,\,j=1}^{L} \la i|h(w)|j \ra d^+_id_j+E_fN_f,
\ee
where the non-vanishing matrix elements are given by
\bea
\label{hxy}
\la i |h(w) |i\ra =Uw_i,\ \ \ \ \ \ \ \la i |h(w) |j\ra =-1 \ \ \ \
\mbox{if} \ \ \ j=i\pm1.
\eea
Let $\{|{\nu}\ra\}_{{\nu}={\nu}_1,\ldots,{\nu}_{L}}$ be the orthonormal basis built
out of the eigenstates of $h(w)$ to the eigenvalues
$\lambda_{\nu}$, such that $\lambda_{\nu} \leq \lambda_{{\nu}'}$ if ${\nu} < {\nu}'$.
Then, the unitary matrix $\cal{U}$, with the following matrix elements
${\cal{U}}_{i{\nu}}$:
\bea
\label{U}
{\cal{U}}_{i{\nu}}=\la i|{\nu}\ra,
\eea
diagonalizes the matrix of $h(w)$,
\bea
\label{UhU}
\sum_{i,j} {\cal{U}}^+_{{\nu}j} \la j|h(w) |i \ra {\cal{U}}_{i{\nu}'}=
\lambda_{\nu}\delta_{{\nu}{\nu}'}.
\eea
Moreover, the set of operators $\{b^+_{\nu}, b_{\nu}\}, \ 
{\nu}={\nu}_1,\ldots,{\nu}_{L}$,
defined by
\bea
\label{b}
b_{\nu}=\sum_{i=1}^{L}\la {\nu}|i\ra d_i ,
\eea
satisfies the canonical anticommutation relations, and
\bea
\label{Hbw}
H(w)=\sum_{i,\,j=1}^{L} \la i|h(w)|j\ra
d^+_id_j+E_fN_f=\sum_{\nu}\lambda_{\nu}b^+_{\nu}b_{\nu}+E_fN_f.
\eea
Since,
\bea
\label{ax}
d_i=\sum_{\nu} \la i|{\nu}\ra b_{\nu},
\eea
we can express the momentum distribution function of itinerant 
electrons
\be
\label{nkk}
n_k=\frac{1}{L}\sum_{j,l}e^{ik(j-l)}\la d^+_jd_l\ra
\ee
in terms of the site-components $\la i|{\nu}\ra$ of the eigenvectors $|{\nu}\ra$:
\be
\label{nk}
n_k=\frac{1}{L}\sum_{j,l}e^{ik(j-l)}\sum_{{\nu}\leq {\nu}_F}\la {\nu}|j\ra \la l|{\nu}\ra, 
\ee
where ${\nu}_F$ stands for the label of eigenvectors $|{\nu}\ra$,
such that there is exactly $N_d$ eigenvectors with ${\nu} \leq {\nu}_F$.
For some low-period ion configurations the eigenproblem can be solved
exactly, while for any other ground-state configurations of interest 
the numerical exact-diagonalization procedure can be used.
In the next section we calculate the momentum distribution of
itinerant electrons for three periodic phases with the smallest periods 
as well as for the phase separated ground states. Such a selection of 
phases is not accidental. In our previous papers~\cite{Fark2,Fark3,EJP} 
we have shown 
that just these phases occupy the largest regions in the ($E_f-U$) 
ground-state phase diagram of the Falicov-Kimball model. Indeed, 
for large $U$ there is only one nontrivial ground-state phase (besides 
the empty or fully occupied phase), and namely the two-period phase 
$w^{(1)}=\{10 \dots 10\}$, and for intermediate values of $U$ there are 
only two other relevant phases and namely the three-period phase  
$w^{(2)}=\{110 \dots 110\}$ and the four-period phase  
$w^{(3)}=\{1110 \dots 1110\}$.  These phases are ground states also
for small values of $U$, but in this region also the phase separated 
configurations, as well as the periodic phases with larger periods can be  
the ground states of the Falicov-Kimball model. The periodic phases are
insulating, and the phase separated configurations are metallic and thus 
one can expect fully different behavior of the momentum distribution in
these ground states. 

\section{Results and discussion}

As was described above one has to know the site-components 
$\la i|{\nu}\ra$ of the eigenvectors $|{\nu}\ra$ to calculate
the momentum distribution of itinerant electrons. In order 
to calculate $\la i|{\nu}\ra$  for $w^{(1)}$, $w^{(2)}$
and $w^{(3)}$ we have generalized the procedure used by Jedrzejewski
et al.~\cite{Jedr} for analytical calculations of correlation functions 
in the two-period phase. 

For $w^{(n)}$, $n=1,2,3$  the nonvanishing matrix elements of 
$h(w^{(n)})$ are given by
\bea
\label{hpch}
\la i|h(w^{(n)})|{i+1}\ra = -1, \ \ \
\la i|h(w^{(n)})|i \ra = Uw^{(n)}_i .
\eea
One can easily verify that the matrix of $h(w^{(n)})$ 
can be rewritten to a block-diagonal form by reordering the original 
basis $\lb |k\ra \rb$, $k=2\pi l/L$, $l=0,1,\ldots, L-1$.
The new basis for $w^{(n)}$, $n=1,2,3$ that reduces
$h(w^{(n)})$ to a block-diagonal form is given by: 
\be
\begin{array}{ll}
\lb |k\ra, |k+\pi\ra \rb, & \ \ \  \  n=1, \nonumber \\  
\lb |k\ra, |k+2\pi/3\ra, |k+4\pi/3\ra  \rb, & \ \ \ \   n=2, \nonumber \\   
\lb |k\ra, |k+\pi/2\ra, |k+\pi\ra, |k+3\pi/2\ra \rb, & \ \ \ \  n=3, \nonumber \\
\end{array} 
\nonumber
\ee
where $k=2\pi l/L$, $l=0,1,\ldots, L/(n+1)-1$.

The diagonal blocks are $n+1 \times n+1$ matrices
$\lbr h(w^{(n)}) \rbr_k$ that in the corresponding basis
given by preceding equation have the form
\be
\label{w1}
\lbr h(w^{(1)}) \rbr_k = \lbr \begin{array}{cc}
\varepsilon_k+\alpha & -\alpha     \\            
             -\alpha & \varepsilon_{k+\pi}+\alpha \\  
\end{array} \rbr, \ \ \ \  \alpha=\frac{U}{2}.
\ee

\be
\label{w2}
\lbr h(w^{(2)}) \rbr_k = \lbr \begin{array}{ccc}
\varepsilon_k+2\alpha & -\alpha  & -\alpha \\
-\alpha & \varepsilon_{k+2\pi/3}+2\alpha & -\alpha \\
-\alpha & -\alpha & \varepsilon_{k+4\pi/3}+2\alpha \\
\end{array} \rbr, \ \ \ \  \alpha=\frac{U}{3}.
\ee

\be
\label{w3}
\lbr h(w^{(3)}) \rbr_k = \lbr \begin{array}{cccc}
\varepsilon_k+3\alpha & -\alpha  & -\alpha & -\alpha \\
-\alpha & \varepsilon_{k+\pi/2}+3\alpha & -\alpha & -\alpha \\
-\alpha & -\alpha & \varepsilon_{k+\pi}+3\alpha & -\alpha \\
-\alpha & -\alpha & -\alpha & \varepsilon_{k+3\pi/2}+3\alpha \\ 
\end{array} \rbr, \ \ \ \  \alpha=\frac{U}{4},
\ee
where $\varepsilon_k=-2\cos k$.

Let, $\lambda^{(n)}_{\mu}(k)$ be eigenvalues of $h(w^{(n)})$, i.e.,
\bea
\label{evevch}
\lbr h(w^{(n)}) \rbr_k
|k\ra^{(n)}_{\mu}=\lambda^{(n)}_{\mu}(k)|k\ra^{(n)}_{\mu}, \ \ \
\mu=1,2,\dots n+1. 
\eea

Then the corresponding eigenvectors $|k\ra^{(n)}_{\mu}$
can be written as:

\be
\begin{array}{l}
\label{vec}
|k\ra^{(1)}_{\mu} = 
x^{(1)}_{\mu}|k\ra +  y^{(1)}_{\mu}|k+\pi\ra,  \nonumber \\  
|k\ra^{(2)}_{\mu} = 
x^{(2)}_{\mu}|k\ra +  y^{(2)}_{\mu}|k+2\pi/3\ra + z^{(2)}_{\mu}|k+4\pi/3\ra,  \nonumber \\   
|k\ra^{(3)}_{\mu} = 
x^{(3)}_{\mu}|k\ra + y^{(3)}_{\mu}|k+\pi/2\ra +  z^{(3)}_{\mu}|k+\pi\ra + v^{(3)}_{\mu}|k+3\pi/2\ra,   \nonumber \\
\end{array} 
\nonumber
\ee
where explicit expressions for coefficients  
$x^{(n)}_{\mu}, y^{(n)}_{\mu}, z^{(n)}_{\mu}, v^{(n)}_{\mu}$, $n=1,2,3$ 
are given in Appendix.

Since we are interested in the half-filled band case $N_d+N_f=L$
(which is the point of the special interest for valence
and metal-insulator transitions caused by promotion of electrons
from localized $f$ orbitals to the conduction
band states) we can restrict our considerations to the lowest $N_d$
eigenvalues $\lambda^{(n)}_{1}(k)$ and the corresponding eigenvectors
$|k\ra^{(n)}_{1}$, and set
\bea
\label{notation}
|k\ra^{(n)} \equiv |k\ra^{(n)}_{1}, \ \ \
x^{(n)}\equiv x^{(n)}_{1}, \ \ \
y^{(n)}\equiv y^{(n)}_{1}, \ \ \
z^{(n)}\equiv z^{(n)}_{1}, \ \ \
v^{(n)}\equiv v^{(n)}_{1}.
\eea
Then, the site-components $\la m|{k}\ra$ for considered phases 
$w^{(n)}, n=1,2,3$ read

\be
\begin{array}{l}
\la m|k\ra^{(1)}=\frac{1}{\sqrt{L}} 
\ld x^{(1)}\mbox{e}^{ikm} +  y^{(1)}\mbox{e}^{i(k+\pi)m}  \rd,   \nonumber \\  
\la m|k\ra^{(2)}=\frac{1}{\sqrt{L}} 
\ld x^{(2)}\mbox{e}^{ikm} +  y^{(2)}\mbox{e}^{i(k+2\pi/3)m} 
+ z^{(2)}\mbox{e}^{i(k+4\pi/3)m} \rd,  \nonumber \\   
\la m|k\ra^{(3)}=\frac{1}{\sqrt{L}} 
\ld x^{(3)}\mbox{e}^{ikm} + y^{(3)}\mbox{e}^{i(k+\pi/2)m} 
+ z^{(3)}\mbox{e}^{i(k+\pi)m} + v^{(3)}\mbox{e}^{i(k+3\pi/2)m} \rd.   \nonumber \\
\end{array} 
\nonumber
\ee

Substituting these expressions into~(\ref{nk}) and doing some tedious 
algebra one obtains the final expression for the momentum distribution 
of itinerant electrons  

\be
\label{nk123}
n^{(i)}_k=|x^{(i)}|^2, \ \ \ \ i=1,2,3.
\ee

Using this expression (and corresponding expressions from Appendix) one can
plot the momentum distribution of itinerant electrons in the particular
phase as a function of $k$ for different values of the Coulomb interaction
$U$. For the two-period phase $w^{(1)}$ the results obtained are displayed
in Fig.~1. It is seen that for all examined $U$ the momentum distribution 
is a smooth function of $k$. There is no sign of any discontinuity at the
Fermi surface $k=k_F=\pi/2$ (the non-Fermi liquid behavior) as well as no 
sign of singular behavior at $k=k_F$ (the non-Luttinger liquid behavior). 
This can be verified analytically since the expression~(\ref{nk123}) 
(for $i=1$) can be rewritten as    

\be
\label{nk1}
n^{(1)}_k=\frac{1}{2}\ld 1- \frac{\varepsilon_k}
{\sqrt{ \varepsilon^2_k+(\frac{U}{2})^2}} \rd.
\ee

Obviously, there is no discontinuity as well as singular behavior at $k_F$
for finite $U$. Near $k_F$, $n^{(1)}_k$ behaves like

\be
\label{nkf}
n^{(1)}_k=\frac{1}{2}\ld 1- \frac{k-\pi/2}
{\sqrt{ (k-\pi/2)^2+U^2}} \rd,
\ee
what corresponds exactly to the behavior of the Tomonoga-Luttinger  
fermions coupled by $2k_F$ potential~\cite{ijmp}. This potential gives rise
to a gap in the energy band spectrum of the Tomonoga-Luttinger 
fermions, and the appearance of this gap results in the smooth
behavior of the momentum distribution. We believe that the same 
mechanism (the existence of a gap in the charge excitation spectrum 
at half-filling) is responsible for a smooth behavior of the momentum
distribution of the Falicov-Kimball model. The same behavior at half-filling
exhibits also the Hubbard model that has a gap in the charge 
excitation spectrum at $n_{\uparrow}=n_{\downarrow} = \frac{1}{2}$,  
but not for any other density. Since the Falicov-Kimball model can be considered 
as an approximation to the full Hubbard model in which one part of electrons 
(say with spin down is immobile), it could be interesting to compare our 
solution~(\ref{nk1}) with known results for the momentum distribution 
of the Hubbard model. 

The analytical results for the momentum distribution $n^H_k$ in the Hubbard 
model are known in the strong coupling limit where the first two terms 
of the perturbation expansion read~\cite{ijmp} 

\be
\label{nkh}
n^{H}_k=\frac{1}{2}\ld 1- \frac{4 \ln 2}{U} \varepsilon_k \rd.
\ee
Comparing this result with the strong coupling expansion   
of~~(\ref{nk1}) 
\be
\label{nkfkm}
n^{H}_k=\frac{1}{2}\ld 1- \frac{2}{U} \varepsilon_k \rd,
\ee
one finds that the Falicov-Kimball model (in spite of the above mentioned 
simplification) contains still much of physics of the full Hubbard model 
at half-filling. This result is not surprising since for both models there 
is the antiferromagnetic long-range order in the ground state for
$n_d=n_f=n_{\uparrow}=n_{\downarrow}=\frac{1}{2}$. 
Since this ground state persists for all Coulomb interactions $U > 0$
(for both the Hubbard and Falicov-Kimball model) one could expect a good 
accordance of results also for intermediate and weak
interactions. For arbitrary $U$ there is
only an approximate (an antiferromagnetic Hartree-Fock) solution for $n^H_k$ 
in the full Hubbard model~\cite{Ogawa}, that has precisely the some form 
as our solution for the Falicov-Kimball model.

Outside the half-filling, the momentum distribution behaves, however fully
differently in the Falicov-Kimball and Hubbard model. While $n_k$ in the
Hubbard model~\cite{Ogata} exhibits the power-low singularity at $k=k_F$ for 
$n_{\uparrow}=n_{\downarrow} \ne \frac{1}{2}$ (even for $U \to \infty$), $n_k$ 
in the Falicov-Kimball model remains still a continuous function of $k$ 
for all finite Coulomb interactions. The situation for 
$n_d=\frac{1}{3}$ (the phase
$w^{(2)}$) is displayed in Fig.~2. A continuous character of $n^{(2)}_k$ is 
obvious for all finite Coulomb interactions $U$, and increasing $U$ only 
smears $n^{(2)}_k$. In accordance with the Hubbard model an unusual behavior 
of the momentum distribution is observed near $n_k=3k_F$ and $U$ small. 
However, while the momentum distribution  has a weak singularity at
$k=3k_F$ in the Hubbard model~\cite{Ogata}, it has a local maximum in the
Falicov-Kimball model at this point. The similar behavior is observed also
for $n_d=\frac{1}{4}$ (see Fig.~3). Again there is a continuous change 
of $n^{(3)}_k$ at $k=k_F$, with no sign of any discontinuity or singular 
behavior for finite $U$. Also an unusual behavior of $n^{(3)}_k$ 
(a local maximum) is observed at $k=3k_F$.                

To reveal the origin of this unusual behavior we have performed the lowest
order perturbation calculations of $n^{(i)}_k$, i=1,2,3 in terms of $U$. 
A straightforward application of the perturbation procedure~\cite{Ogata} 
yields the following expression for the momentum distribution of itinerant 
electrons in the Falicov-Kimball model (up to the second order):  

\be
\label{nk_pert}
n^{(l)}_k=U^2\sum_{k'} \frac{|V^{(l)}_{k,k'}|^2}
{\ld \varepsilon_k-\varepsilon_{k'} \rd^2}n^{(0)}_{k'}, \ \ \   
\mbox{for} \ \ |k| > k_F,  \ \ \ (l=1,2,3),
\ee
where
\be
\label{Vkk}
V^{(l)}_{k,k'}=\frac{1}{L}\sum_{j}\mbox{e}^{i(k-k')R_j}w^{(l)}_j 
\ee
and
$n^{(0)}_k$ is the Fermi distribution function of noninteracting electrons.

For the periodic phases $w^{(l)}$ (l=1,2,3) the matrix elements 
$V^{(l)}_{k,k'}$ can be directly calculated and the off-diagonal 
elements, that enter to Eq.~(\ref{nk_pert}), are given by
\be
V^{(l)}_{k,k'}=\left \{ \begin{array}{ll}
   -\frac{1}{l+1},   & \mbox{for}  \ \ k'=k \pm 2nk_F, \ \ n=1,2 \dots l, \\
  \quad       0,     &   \mbox{otherwise.}   \\
\end{array}
\right.
\ee
Substituting this expression into Eq.~(\ref{nk_pert}) one obtains the 
final expression for $n^{(l)}_k$
\be
\label{nk_pert2}
n^{(l)}_k=\frac{1}{(l+1)^2}\frac{U^2}
{(\varepsilon_k-\varepsilon_{k - 2nk_F})^2 },
\ee
for $k_F(1 + 2(n-1)) < k < k_F(1 + 2n)$. 

Analysing this expression one can 
find that $n^{(l)}_k$ changes its behavior 
at points $k=3k_F, 5k_F, 7k_F, \dots$ and this change can produce 
the local maxima at some points. Indeed, we have found that the second order
perturbation theory can describe the local maximum at $k=3k_F$ for 
$n_d=\frac{1}{3}$. This is illustrated in Fig.~4 where the momentum 
distribution (exact as well as perturbation results) 
is displayed for the periodic phases with the period three, four 
and six. It is clear that this unusual behavior of the momentum
distribution at $k=3k_F, 5k_F, 7k_F \dots$ is caused by the 
periodic arrangement of the localized $f$ electrons, which results in 
a very simple form of off-diagonal matrix elements $V^{(l)}_{k,k'}$. 
It is interesting that a similar unusual behavior of the momentum 
distribution at $k= 3k_F, 5k_F, 7k_F \dots$ has been observed 
(predicted) also in the Hubbard model, however in this case it was assigned 
to a pair (multipairs) of electron-hole excitations~\cite{Ogata}. 

Finally, let us briefly discuss the case of phase separation. It is
well-known~\cite{Fark3} that the ground states of the Falicov-Kimball model for
sufficiently small Coulomb interactions ($U<1$) and sufficiently small 
($n_d<1/4$) or large ($3/4<n_d<1$) $d$-electron concentrations
are the phase separated configurations, i.e., configurations in
which one-half of lattice is fully empty or fully occupied by $f$-electrons. 
Such configurations are metallic and we
expect fully different behavior of the momentum distribution of itinerant
electrons in this phase. Unfortunately, the phase separated configurations
are not periodic and thus we cannot proceed in the analytic calculations as in
the preceding cases, however, the  numerical procedure is still possible. To
calculate $n_k$ numerically in the phase separated region ($U$ and $n_d$
small)  we need to know exactly the ground-state configuration $w^{ps}$
for selected $U$ and $n_d$. In general, it is very difficult to find the
ground-state configuration for arbitrary $U$ and $n_d$ from the phase
separated region, however, it is possible to solve this task for some special values
of $U$ and $n_d$. For example, exhaustive small-cluster exact-diagonalization
studies that we have performed in our previous paper~\cite{Fark3} for 
the Falicov-Kimball model in the weak-coupling limit showed that the
ground-state configuration of the model for $U=0.6$ and $n_d=1/8$ is of the
type $w^{ps}=\{111100\dots111100111\dots111\}$. The momentum distribution
$n^{ps}_k$ for this configuration (calculated numerically using the
procedure described in the preceding section) is displayed in Fig.~5a. As we
conjectured the momentum distribution of itinerant electrons behaves fully
differently in the metallic phase. It seems that there is a discontinuity
(or a singular behavior) in $n^{ps}_k$ at some critical value of
$k=k_0<k_F$. To determine exactly which type of behavior realizes near $k_0$
we have calculated the finite-size discontinuity 
$\Delta =n^{ps}_{k_0+\frac{2\pi}{L}}-n^{ps}_{k_0-\frac{2\pi}{L}}$
as a function of $1/L$. The results obtained are plotted in
Fig.~5b and they clearly show that there is a discontinuity at $k=k_0$.
This shows on the Fermi liquid behaviour of itinerant electrons
(or some part of them) in the phase separated (metallic) state. 
To verify this conjecture we have calculated separately contributions 
to $n^{ps}_k$ from different parts of a lattice. Since the configuration   
$w^{ps}$ can be formally considered as a mixture of two phases    
$w^{a}=\{111100\dots111100 \}$ and $w^{b}=\{111\dots111\}$
it is natural to divide the lattice into two parts $a$ and $b$ 
and to rewrite the summations in  Eq.~(\ref{nk}) as follows
\bea
\label{nkab}
n^{ps}_k &=&\frac{1}{L}\sum_{j,l}e^{ik(j-l)}\sum_{{\nu}\leq {\nu}_F}\la 
{\nu}|j\ra \la l|{\nu}\ra=\sum_{j,l}\Omega_{j,l} = 
n^{a}_k+n^{ab}_k+n^{ba}_k+n^{b}_k \non
   &=& 
\sum_{j\in a,l\in a}\Omega_{j,l}+
\sum_{j\in a,l\in b}\Omega_{j,l}+
\sum_{j\in b,l\in a}\Omega_{j,l}+
\sum_{j\in b,l\in b}\Omega_{j,l}. 
\eea

The momentum dependence of single contributions 
$n^{a}_k, n^{ab}_k, n^{ba}_k$ and $n^{b}_k$
is plotted in Fig.~6a for two values of $L$. 
It is seen that with increasing $L$ the contribution
$n^{ab}_k$ + $n^{ba}_k$ goes to zero and thus  
in the thermodynamic limit $L\to \infty$ only $n^{a}_k$ and $n^{b}_k$
remain finite. The first contribution $n^{a}_k$ exhibits
the same behavior as the momentum distribution in the periodic
(insulating) phases $w^{(1)}$, $w^{(2)}$ and $w^{(3)}$. There
is no sign of any discontinuity or singular behavior at $k=k_F$.
This result is expected since also the phase $w^{a}$ is periodic.
In contrary to this case the second contribution $n^{b}_k$ 
exhibits an unexpected behavior of the Fermi liquid type with
discontinuity at some critical momentum $k=k_0=0.1\pi$ that does not
coincides however with the Fermi surface momentum 
$k_F=\frac{N_d}{L}\pi=0.125\pi$. Such a behavior as well as the meaning
of $k_0$ can be easily explained within the formalism described above
(a mixture of two phases). Let $N^{b}_d$ be the number of itinerant 
electrons in the phase $w^{b}$. It can be calculated directly from 
the expression for $n_k$ (Eq.~(\ref{nk})) putting $j=l$ and taking
the sum only over the lattice sites from $w^{b}$, i.e,   
\be
\label{Nd}
N^{b}_d=\sum_{j\in b}\sum_{{\nu}\leq {\nu}_F}\la {\nu}|j\ra \la j|{\nu}\ra. 
\ee

In Fig.~6b we plotted the quantity  
$k^{b}_F=\frac{N^{b}_d}{L^b}\pi$, that represents the Fermi surface 
of itinerant electrons in $w^{b}$, as a function of $1/L$ 
($L^{b}=5L/8$ denotes the size of $w^{b}$ phase). It is seen that 
$k^{b}_F$ goes to $k_0$ for $L \to \infty$ what provides a clear physical 
interpretation for $k_0$.  According these results the itinerant electrons in 
the phase $w^{b}$ behaves like the Fermi liquid with a discontinuity at 
$k^{b}_F=\frac{N^{b}_d}{L^{b}}\pi=k_0$.

In summary, the momentum distribution $n_k$ of itinerant electrons in 
the one-dimensional Falicov-Kimball model has been calculated  
for various ground-state phases. 
In particular, we have examined the periodic phases with period two, 
three and four (that are ground-states for all Coulomb interactions) as well 
as the phase separated states (that are ground states for small Coulomb 
interactions).  We have found that for all periodic phases examined 
the momentum distribution is a smooth function of $k$ with no sign of 
any discontinuity or singular behavior at the Fermi surface $k=k_F$. 
An unusual behavior of $n_k$ (a local maximum) is found at $k=3k_F$ 
for electron concentrations outside half-filling. For the phase separated 
ground states the momentum distribution $n_k$ exhibits discontinuity 
at $k=k_0 < k_F$. This behavior has been interpreted in terms 
of a Fermi liquid.  

\vspace{0.3cm}
\noindent
{\bf Acknowledgements}
\\[3mm]
This work was supported by the Science and Technology 
Assistance Agency under Grant APVT-51-021602.
Numerical results were obtained using
the PC-Farm of the Slovak Aca\-de\-my of Sciences.

\section{Appendix}
Here we give the explicit expressions for coefficients  
$x^{(n)}_{\mu}, y^{(n)}_{\mu}, z^{(n)}_{\mu}, v^{(n)}_{\mu}$ 
from~(\ref{vec}) and the explicit expressions for eigenvalues
$\lambda^{(n)}_{\mu}(k)$ of $h(w^{(n)})$, ($n=1,2,3$):

\be
\begin{array}{l}
x^{(1)}_{\mu} = 
\frac{1-b^{(1)}_{\mu}}{1-a^{(1)}_{\mu}}g^{(1)}_{\mu},  \ \ \  
y^{(1)}_{\mu} = g^{(1)}_{\mu},  \nonumber \\ 
g^{(1)}_{\mu} = 
\lb 1
+\ld \frac{1-b^{(1)}_{\mu}}{1-a^{(1)}_{\mu}} \rd^2
\rb^{-\frac{1}{2}},    
\end{array} 
\ee

\be
\begin{array}{l}
x^{(2)}_{\mu} = 
\frac{1-c^{(2)}_{\mu}}{1-a^{(2)}_{\mu}}g^{(2)}_{\mu},  \ \ \  
y^{(2)}_{\mu} = 
\frac{1-c^{(2)}_{\mu}}{1-b^{(2)}_{\mu}}g^{(2)}_{\mu},  \ \ \  
z^{(2)}_{\mu} = g^{(2)}_{\mu},  \nonumber \\
g^{(2)}_{\mu} = 
\lb 1
+\ld \frac{1-c^{(2)}_{\mu}}{1-a^{(2)}_{\mu}} \rd^2
+\ld \frac{1-c^{(2)}_{\mu}}{1-b^{(2)}_{\mu}} \rd^2
\rb^{-\frac{1}{2}},  \nonumber \\   
\end{array} 
\ee

\be
\begin{array}{l}
x^{(3)}_{\mu} = 
\frac{1-d^{(3)}_{\mu}}{1-a^{(3)}_{\mu}}g^{(3)}_{\mu},  \ \ \  
y^{(3)}_{\mu} = 
\frac{1-d^{(3)}_{\mu}}{1-b^{(3)}_{\mu}}g^{(3)}_{\mu},  \ \ \  
z^{(3)}_{\mu} = 
\frac{1-d^{(3)}_{\mu}}{1-c^{(3)}_{\mu}}g^{(3)}_{\mu},  \ \ \  
v^{(3)}_{\mu} = g^{(3)}_{\mu},  \nonumber \\
g^{(3)}_{\mu} = 
\lb 1
+\ld \frac{1-d^{(3)}_{\mu}}{1-a^{(3)}_{\mu}} \rd^2
+\ld \frac{1-d^{(3)}_{\mu}}{1-b^{(3)}_{\mu}} \rd^2
+\ld \frac{1-d^{(3)}_{\mu}}{1-c^{(3)}_{\mu}} \rd^2
\rb^{-\frac{1}{2}},  \nonumber \\  
\end{array} 
\ee
where
\be
\begin{array}{l}
a^{(1)}_{\mu}(k)=-\frac{2}{U}\varepsilon_k-1+\frac{2}{U} \lambda^{(1)}_{\mu}(k),
\nonumber \\
b^{(1)}_{\mu}(k)=a^{(1)}_{\mu}(k+\pi), \nonumber 
\end{array} 
\ee

\be
\begin{array}{l}
a^{(2)}_{\mu}(k)=-\frac{3}{U}\varepsilon_k-2+\frac{3}{U} \lambda^{(2)}_{\mu}(k), 
\nonumber \\
b^{(2)}_{\mu}(k)=a^{(2)}_{\mu}(k+2\pi/3), \ \ \
c^{(2)}_{\mu}(k)=a^{(2)}_{\mu}(k+4\pi/3), \nonumber   
\end{array} 
\ee

\be
\begin{array}{l}
a^{(3)}_{\mu}(k)=-\frac{4}{U}\varepsilon_k-3+\frac{4}{U} \lambda^{(3)}_{\mu}(k), 
\nonumber \\ 
b^{(3)}_{\mu}(k)=a^{(3)}_{\mu}(k+\pi/2), \ \ \
c^{(3)}_{\mu}(k)=a^{(3)}_{\mu}(k+\pi), \ \ \
d^{(3)}_{\mu}(k)=a^{(3)}_{\mu}(k+3\pi/2), \nonumber  
\end{array} 
\ee
and
\be
\begin{array}{l}
\lambda^{(1)}_{1}(k)=\frac{U}{2}-\sqrt{(\frac{U}{2})^2+ \varepsilon^2_k},   \nonumber \\
\lambda^{(1)}_{2}(k)=\frac{U}{2}+\sqrt{(\frac{U}{2})^2+ \varepsilon^2_k},   \nonumber \\
\end{array} 
\ee

\be
\begin{array}{l}
\lambda^{(2)}_{1}(k)=-2r\cos(\frac{\phi}{3})+\frac{2U}{3}   \nonumber \\
\lambda^{(2)}_{2}(k)=2r\cos(\frac{\pi}{3}+\frac{\phi}{3})+\frac{2U}{3}  \nonumber \\
\lambda^{(2)}_{3}(k)=2r\cos(\frac{\pi}{3}-\frac{\phi}{3})+\frac{2U}{3}  \nonumber \\
q=cos(3k)+(\frac{U}{3})^3, \ \ \ r=\sqrt{(\frac{U}{3})^2+1}, \ \ \
\phi=\arccos \ld \frac{q}{r^3} \rd, \nonumber
\end{array} 
\ee

\be
\begin{array}{l}
\lambda^{(3)}_{1}(k)=-\frac{\sqrt{6}}{12}p-\frac{1}{12}
(48a-6q-288cq^{-1}-24a^2q^{-1}+72\sqrt{6}bp^{-1})^{1/2}+\frac{3}{4}U,
\nonumber \\
\lambda^{(3)}_{2}(k)=-\frac{\sqrt{6}}{12}p+\frac{1}{12}
(48a-6q-288cq^{-1}-24a^2q^{-1}+72\sqrt{6}bp^{-1})^{1/2}+\frac{3}{4}U,
\nonumber \\
\lambda^{(3)}_{3}(k)=\frac{\sqrt{6}}{12}p-\frac{1}{12}
(48a-6q-288cq^{-1}-24a^2q^{-1}-72\sqrt{6}bp^{-1})^{1/2}+\frac{3}{4}U,
\nonumber \\
\lambda^{(3)}_{4}(k)=\frac{\sqrt{6}}{12}p+\frac{1}{12}
(48a-6q-288cq^{-1}-24a^2q^{-1}-72\sqrt{6}bp^{-1})^{1/2}+\frac{3}{4}U,
\nonumber \\
\end{array} 
\ee

\be
\begin{array}{l}
a=3U^2/8+4, \ \ \ b=U^3/8, \ \ \ c=U^2/4-3U^4/256-2\cos(4k)+2, \nonumber \\
p=(4a+q+48cq^{-1}+4a^2q^{-1})^{1/2}, \nonumber \\  
q=(288ac+108b^2-8a^3+12s)^{1/3}, \nonumber \\
s=(-768c^3+384c^2a^2-48ca^4+432acb^2+81b^4-12b^2a^3)^{1/2}. \nonumber
\end{array} 
\ee

\newpage

\centerline{\bf Figure Captions}

\vspace{0.5cm}
Fig.~1. The momentum distribution of the one dimensional 
Falicov-Kimball model calculated for $n_d=\frac{1}{2}$ and several
different values of $U$.

\vspace{0.5cm}
Fig.~2. (a) The momentum distribution of the one dimensional 
Falicov-Kimball model calculated for $n_d=\frac{1}{3}$ and several
different values of $U$.
(b) The weak-coupling results for the momentum distribution on 
the enlarged scale.

\vspace{0.5cm}
Fig.~3. (a) The momentum distribution $n_k$ of the one dimensional 
Falicov-Kimball model calculated for $n_d=\frac{1}{4}$ and several
different values of $U$.
(b) The weak-coupling results for the momentum distribution on 
the enlarged scale.

\vspace{0.5cm}
Fig.~4. (a) The lowest-order perturbation results for the momentum 
distribution of the Falicov-Kimball model calculated for three different 
$d$-electron concentrations.
(b) A comparison of exact and perturbation results obtained for $U=0.1$
and two different $d$-electron concentrations.

\vspace{0.5cm}
Fig.~5. (a) The momentum distribution $n^{ps}_k$ of the one dimensional 
Falicov-Kimball model calculated numerically for $n_d=\frac{1}{8}$, 
$U=0.6$ and $L=3200$ sites.
(b) The finite-size discontinuity 
$\Delta =n^{ps}_{k_0+\frac{2\pi}{L}}-n^{ps}_{k_0-\frac{2\pi}{L}}$
as a function of $1/L$.

\vspace{0.5cm}
Fig.~6. (a) Contributions to the momentum distribution $n^{ps}_k$ of the 
one dimensional Falicov-Kimball model at $n_d=\frac{1}{8}$ from  
different parts of lattice ($U=0.6$ and $L=3200$).
(b) $k^{b}_F=\frac{N^{b}_d}{L^b}\pi$
as a function of $1/L$.

\newpage
\begin{figure}[htb]
\hspace{-2cm}
\includegraphics[angle=0,width=18.0cm,scale=1]{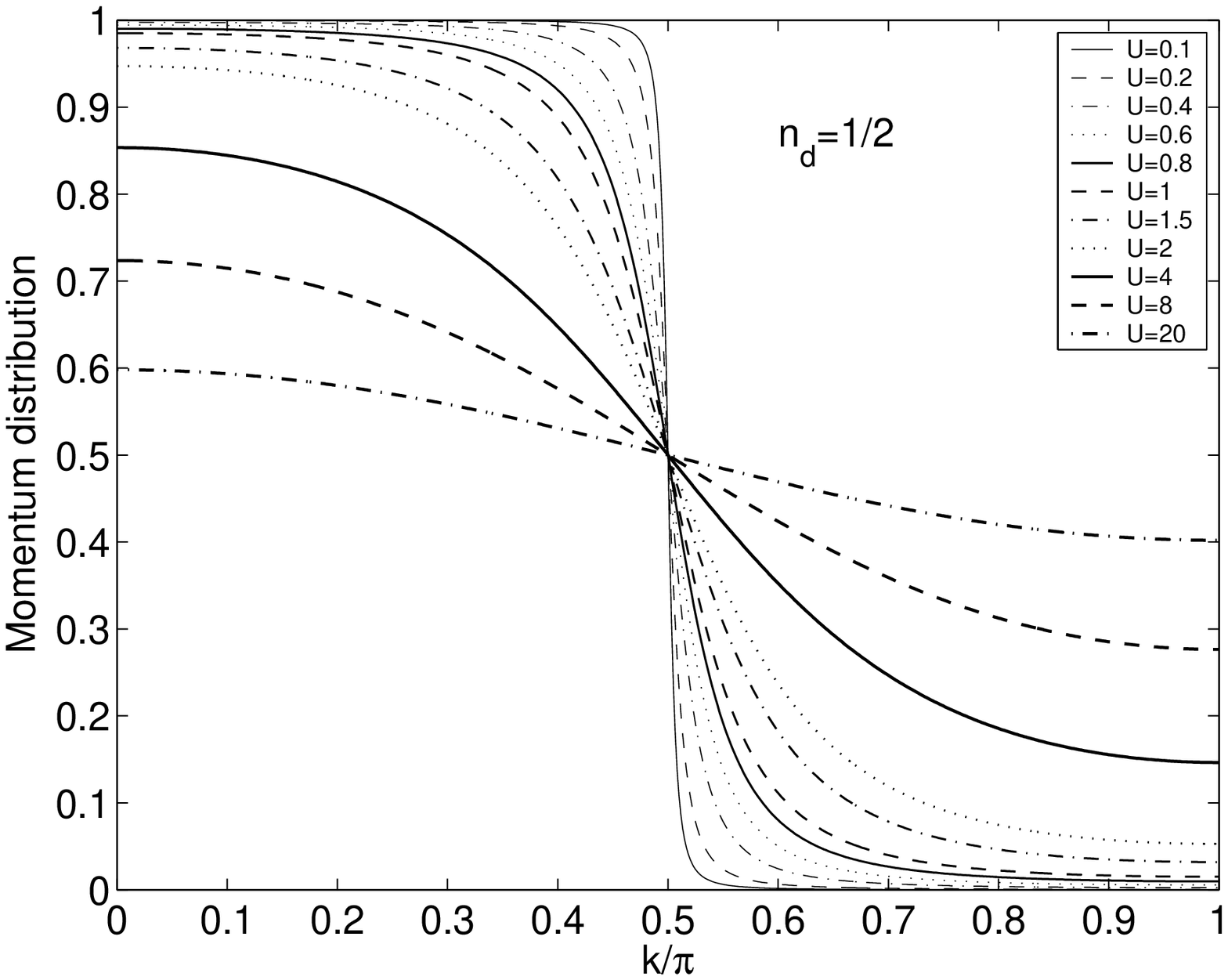}
\caption{ }
\label{fig1}
\end{figure}

\newpage
\begin{figure}[htb]
\hspace{-2cm}
\includegraphics[angle=0,width=18.0cm,scale=1]{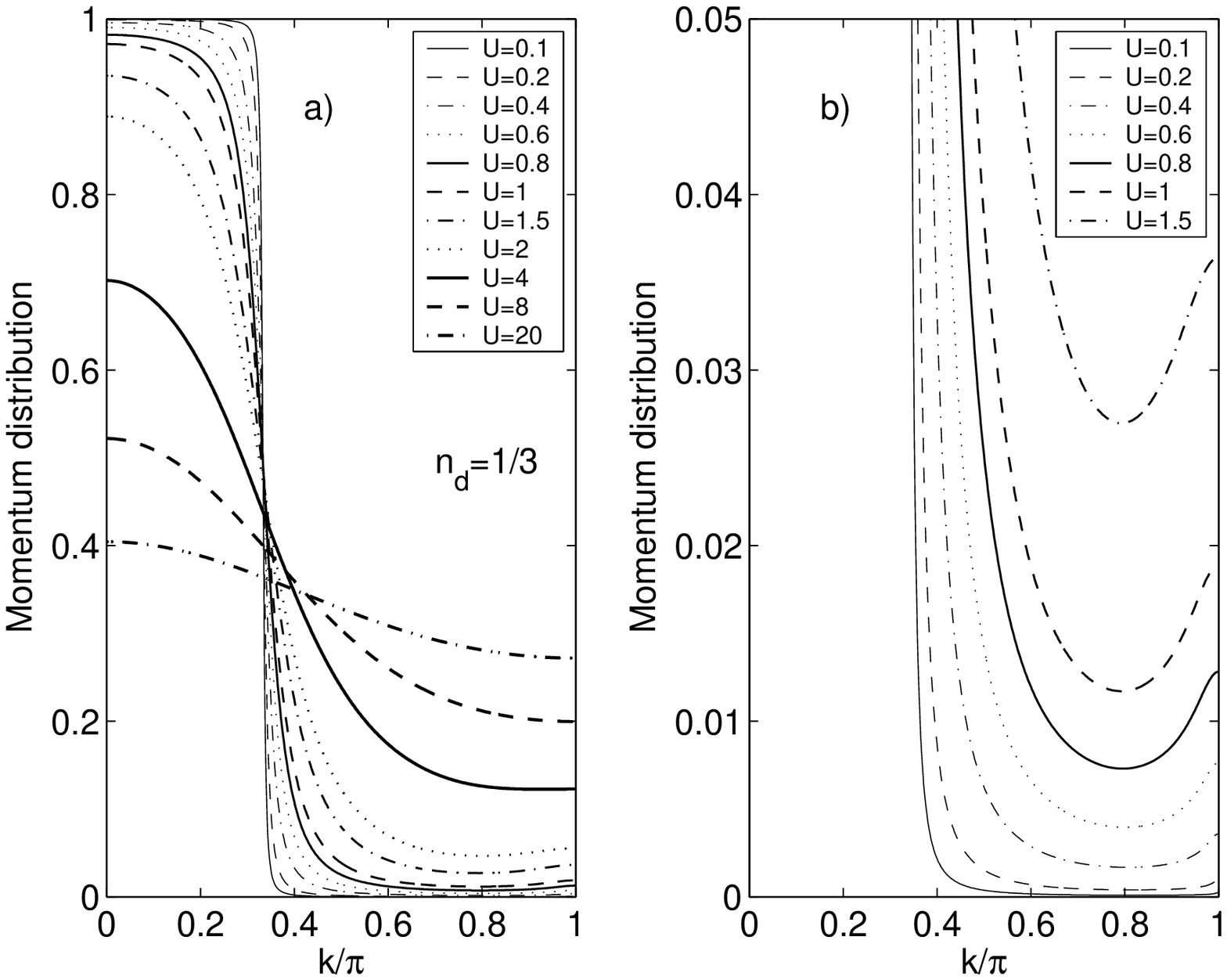}
\caption{ }
\label{fig2}
\end{figure}

\newpage
\begin{figure}[htb]
\hspace{-2cm}
\includegraphics[angle=0,width=18.0cm,scale=1]{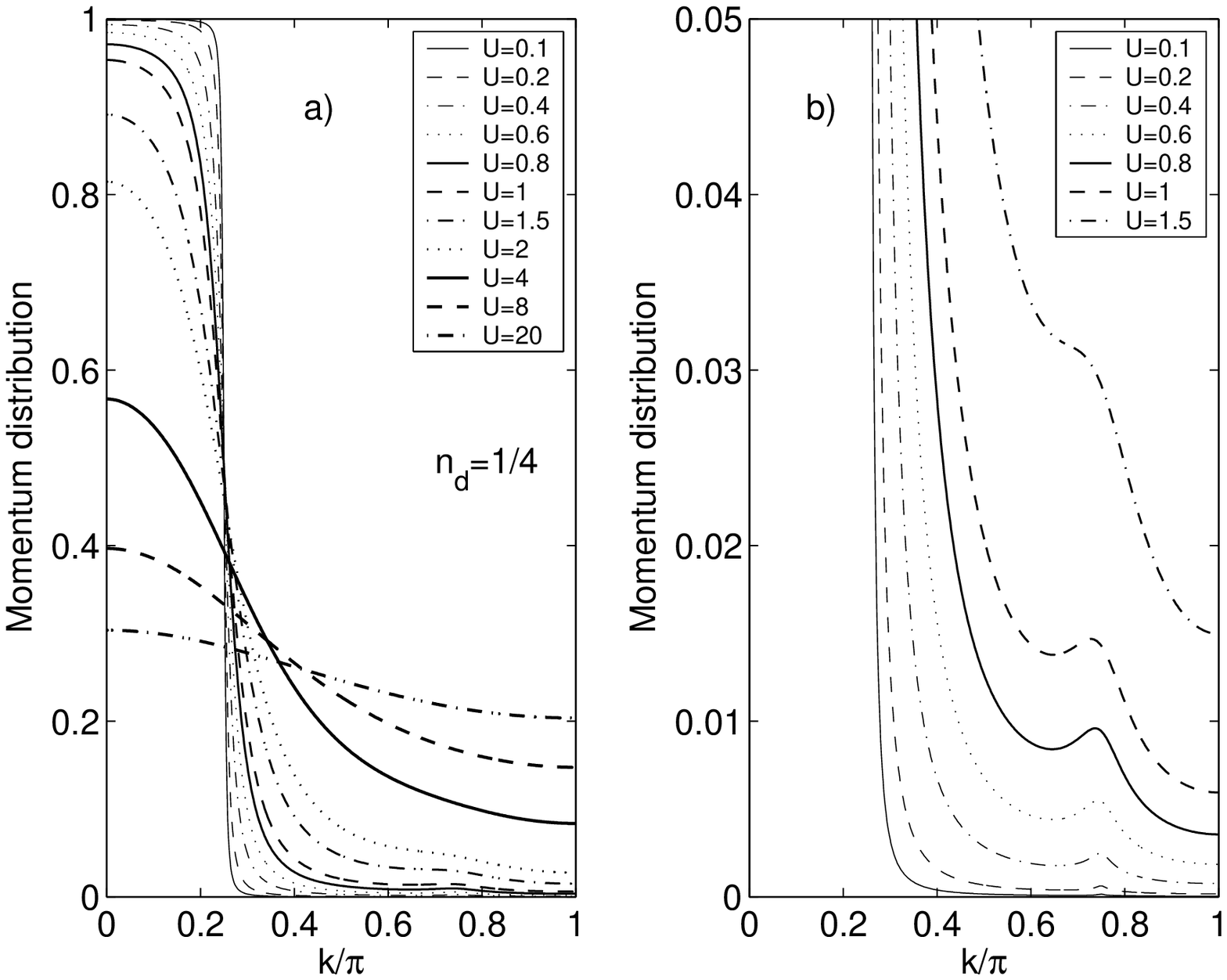}
\caption{ }
\label{fig3}
\end{figure}

\newpage
\begin{figure}[htb]
\hspace{-2cm}
\includegraphics[angle=0,width=18.0cm,scale=1]{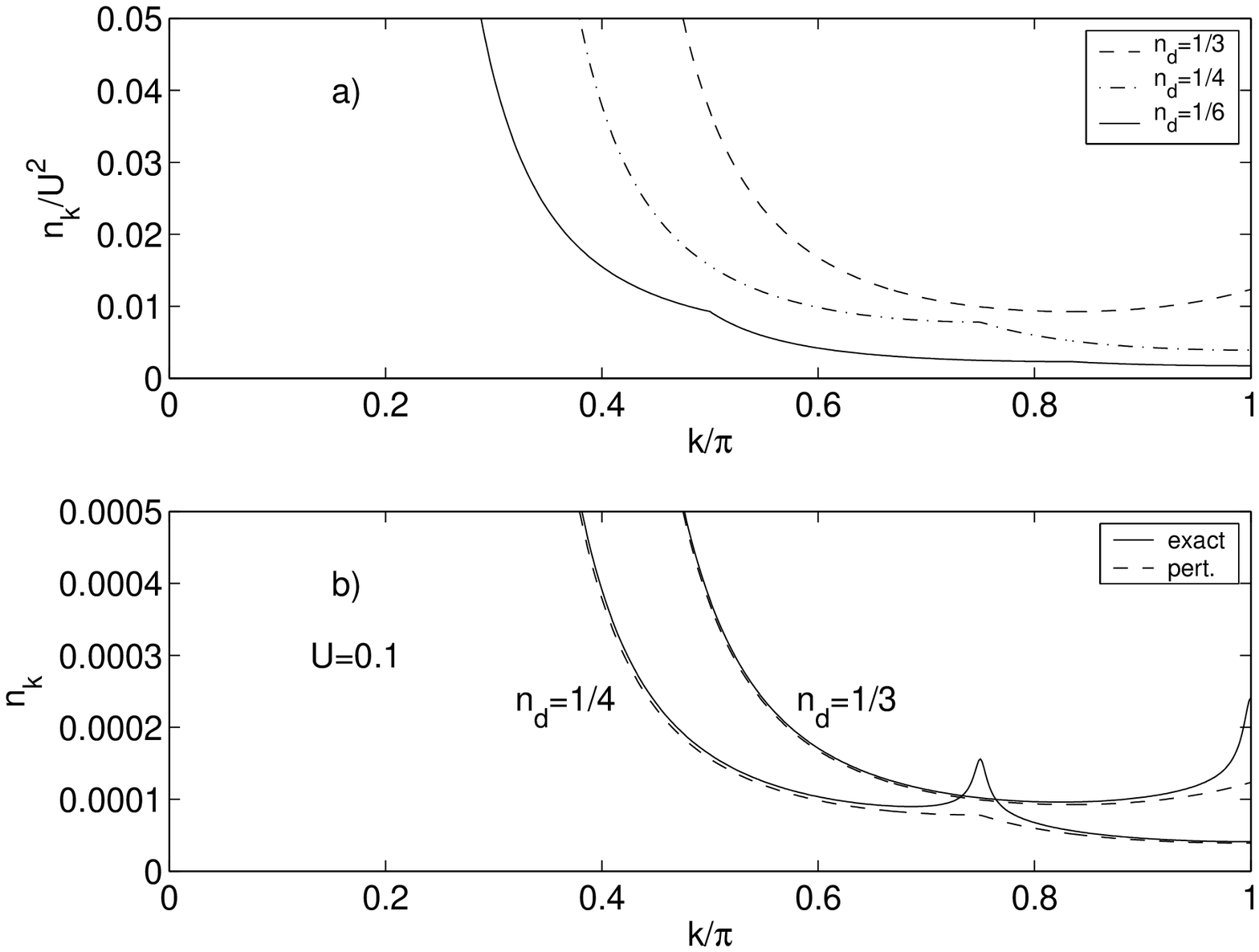}
\caption{ }
\label{fig4}
\end{figure}

\newpage
\begin{figure}[htb]
\hspace{-2cm}
\includegraphics[angle=0,width=18.0cm,scale=1]{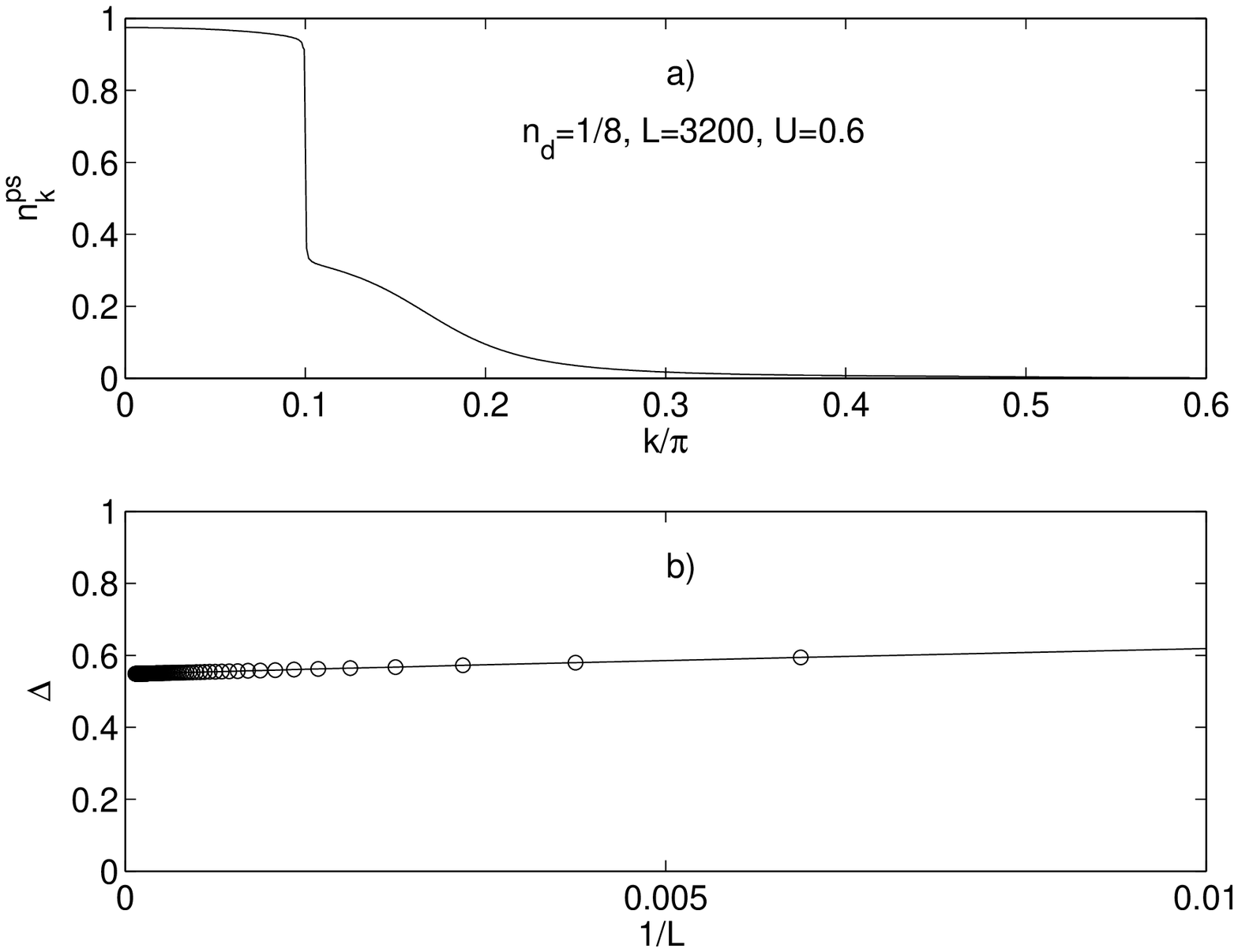}
\caption{ }
\label{fig5}
\end{figure}

\newpage
\begin{figure}[htb]
\hspace{-2cm}
\includegraphics[angle=0,width=18.0cm,scale=1]{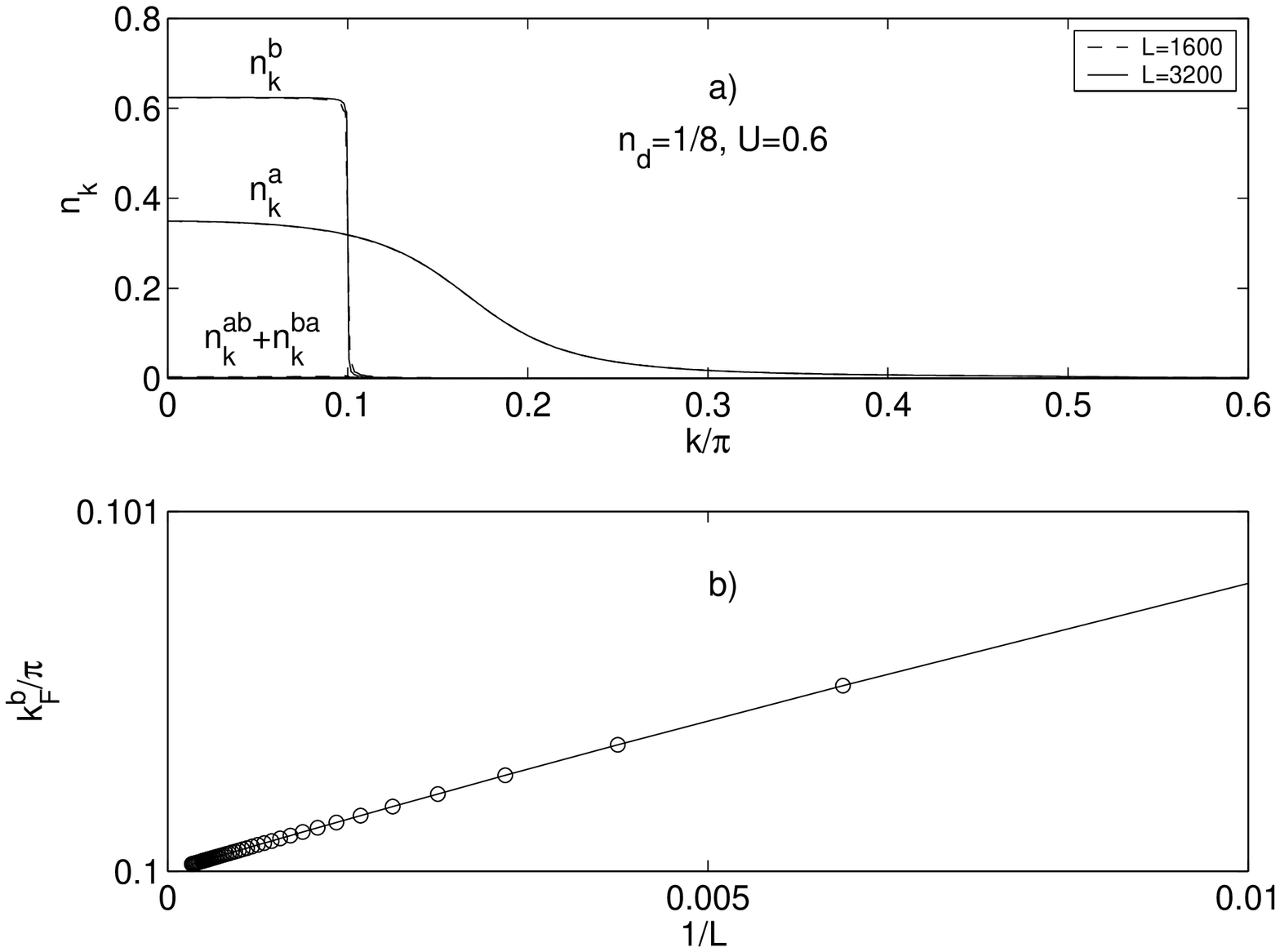}
\caption{ }
\label{fig6}
\end{figure}

\end{document}